# Plasmonic-enhanced photoluminescence in porous silicon with pore-embedded gold nanoparticles fabricated by direct reduction of chloroauric acid.

Salvador Escobar[1,*], Rocío Nava[2], J. A. Reyes-Esqueda[1,3,*].

1 Instituto de Física, Universidad Nacional Autónoma de México, AP 20-364, 01000 CDMX, México

2 Instituto de Energías Renovables, Universidad Nacional Autónoma de México, Privada Xochicalco s/n, Temixco, Morelos, C.P. 62580, México.

3 Département de Physique, Faculté des sciences, Université de Sherbrooke, J1K 2R1, Sherbrooke, QC, Canada

* Corresponding authors: seguerrero@fisica.unam.mx, reyes@fisica.unam.mx

**Abstract**

The low efficiency of porous silicon (p-Si) luminescence hinders the development of silicon-based optoelectronic devices. The increase in p-Si emission using near-field enhancement, owing to the incorporation of gold nanoparticles (AuNPs) into the photonic structure, is probably the most viable alternative. However, the coupling of plasmon resonance to p-Si emission is challenging because of the difficulty in controlling the size and location of the AuNPs with respect to the emissive p-Si layer. In this study, AuNPs were synthesized by clean direct reduction of chloroauric acid inside a p-Si photonic structure. As a result, AuNPs could be synthesized all along the pores of the p-Si structure, allowing to obtain a six-fold enhancement of the p-Si photoluminescence, specifically for the emission band at 567 nm owing to the plasmon effect. Possible applications of this hybrid material include light-emitting devices and photoluminescence-based sensors.

## 1. Introduction

The discovery of visible photoluminescence (PL) of p-Si at room temperature in 1991 led to the development of Si-based optoelectronic devices [1]. However, to date, its low emission efficiency and long lifetime have been obstacles for this purpose. Luminescent porous silicon is usually produced by the electrochemical etching of low-doped crystalline silicon (c-Si) wafer (1-10 Ω-cm) and has been mainly used to develop LEDs with efficiencies below 1% [2-5]. Additionally, this procedure allows the tuning of the p-Si refractive index over a wide range, and the fabrication of 1D photonic structures. Its large surface area has been exploited as a controlled drug delivery vehicle and for the development of highly sensitive sensors [6-9].

In recent years, the surface plasmon resonance of gold nanoparticles (AuNPs) on p-Si has been studied to enhance the luminescence [10,11], sensing [12-14], solar cell efficiency [15,16], and Surface Enhanced Raman Spectroscopy (SERS) [17,18]. Plasmons produce a strongly localized electric field near the surface of nanostructured metals in response to electromagnetic waves [11,19]. When the frequency of the AuNPs surface plasmon resonates with p-Si emission, new radiative recombination channels are induced, decreasing the lifetime, and enhancing the emission intensity [19,20]. AuNPs can be synthesized on p-Si using different methods involving the reduction of chloroauric acid ($HAuCl_4$). Among them are the Turkevitch method, which uses sodium citrate as a reducing and stabilizing agent [11,21]; the Brust two-phase method requires thiol ligands as a capping and reducing agent in the first phase, followed by sodium borohydride as the reducing agent in the second phase [22,23]. These methods have been applied to produce colloidal AuNPs, which are then synthesized on the p-Si surface using different techniques, such as drop casting or immersion, to improve the p-Si luminescence by plasmonic effects [10,11]. In situ methods can be applied to grow AuNPs into porous silicon by direct reduction of $HAuCl_4$ [17]. Different reactants and conditions were used to reduce the gold ions in chloroauric acid. Cetyltrimethylammonium bromide or hydrofluoric acid can be added to promote chloroauric acid reduction [14]. It has also been reported that chloroauric acid is spontaneously reduced in hydrogenated p-Si, and sonication facilitates the growth of AuNPs into p-Si [17]; however, p-Si samples with high porosity are not viable.

In this work, we produced p-Si by electrochemical etching of p-type c-Si wafers with low resistivity (0.01-0.02 Ω-cm) as a platform for the future development of active photonic structures with plasmon surface resonance. AuNPs were synthesized directly in p-Si by in-situ reduction of chloroauric acid, without any additional reducing agent, to avoid contamination. We found that it is necessary to partially oxidize p-Si to grow AuNPs homogeneously inside the porous layer and to avoid their accumulation on the surface. The structure of p-Si and the effect of AuNPs plasmon on luminescence were analyzed. The plasmonic effect of AuNPs in p-Si was detected by Raman and photoluminescence measurements, which showed a six-fold increase in the light emission.

## 2. Materials and Methods

### 2.1 Sample preparation

P-Si was produced by electrochemical etching of boron-doped p-type c-Si wafers with a crystallographic orientation (100) and electrical resistivity of 0.01-0.02 Ω-cm. An aluminum film was deposited on the backside of the c-Si wafers and annealed at 500 °C for 30 min in a $N_2$ atmosphere to make electrical contact. An electrolyte composed of ethanol, hydrofluoric acid, glycerin, and a POM solution in a 6:3:1:1 volumetric ratio was used to produce single p-Si layers. The POM solution was composed of phosphomolybdic acid ($H_3Mo_{12}O_{40}P$) at a concentration of $10^{-3}$ M in hydrogen peroxide and ethanol in a 1:3 volumetric ratio. A current density of 100 mA cm$^{-2}$ for 83 s was applied to produce a thick p-Si layer of approximately 3 µm and porosity of approximately 90%, as determined by gravimetry. At the end of the electrochemical etching, the samples were rinsed several times with ethanol to remove electrolyte residuals. The p-Si samples were partially oxidized by immersion in a solution of hydrogen peroxide and ethanol at a 1:1 volumetric ratio for 24 h and then rinsed with ethanol. AuNPs were synthesized by immersing p-Si samples in a solution prepared by diluting 0.5 mL of aqueous chloroauric acid, with a concentration of 6.25×10$^{-5}$ M, in 30 mL of ethanol. To investigate the effect of p-Si oxidation on the synthesis of AuNPs, seven p-Si samples on the c-Si substrate were prepared: one was kept as reference, one was partially oxidized, and another one, immediately after etching (without oxidation), was immersed in the HAuCl$_4$ solution for 60 s, rinsed with ethanol, and finally oxidized in the hydrogen peroxide solution for 24 h. The rest of the samples were first oxidized in the hydrogen peroxide solution during 24 hours, then immersed in the HAuCl$_4$, varying the dipping time from 30, 60, 90 and 120 minutes. A second set of five p-Si samples was detached from the c-Si substrate by electropolishing at a current density of 400 mA cm$^{-2}$. They were then partially oxidized and immersed in the HAuCl$_4$ to synthesize AuNPs in situ. This set of samples was placed on a microscope cover glass for transmittance measurements.

**2.2 Characterization**

The morphology and elemental composition of the samples were examined by energy-dispersive X-ray spectroscopy (EDS) and SEM images were obtained using a field-emission scanning electron microscope (Hitachi FE-SEM, model S-5500). FTIR spectra were recorded in the range from 400 cm$^{-1}$ to 4000 cm$^{-1}$ using Thermo Scientific equipment (model Nicolet iS 50). Transmittance spectra were measured using a UV-Vis/NIR Spectrophotometer (Jasco model V-670), and Raman spectra were obtained using Oxford Instruments (WITec model alpha300 R). The optical setup used to measure the PL of the samples is illustrated in Fig. 1. An ultraviolet LED (OSRAM model LZ4-V4UV0R-0000), with maximum emission at 365 nm and power at a constant current, was used as the excitation source. The LED emission was focused with a plano-convex lens (Thorlabs, model LA1102-A), projecting a square footprint of 2 mm on the sample. The sample was placed on a height-adjustable swivel positioner (Newport model M-488), and its emission was collected using an optical fiber (Edmund Optics model 58-458) at 1.8 cm and an angle of 67°. The PL spectra were digitized using a spectrometer (BWTEK model Exemplar). Each spectrum was recorded with an integration time of 1 second and averaging 5 measurements. Excitation spectra were measured using an EKSPLA PL2231-50-SH/TH Nd:YAG pulsed laser system featuring ~26 ps pulses with a repetition rate of 50 Hz coupled with an EKSPLA PG 401/SH Optical Parametric Generator. The beam impinged normal to the sample, while the emission was collected at 45° using an optical fiber (Ocean Optics model P1000-2-UV-VIS), and analyzed using an Ocean Optics USB2000+ UV-VIS spectrometer.

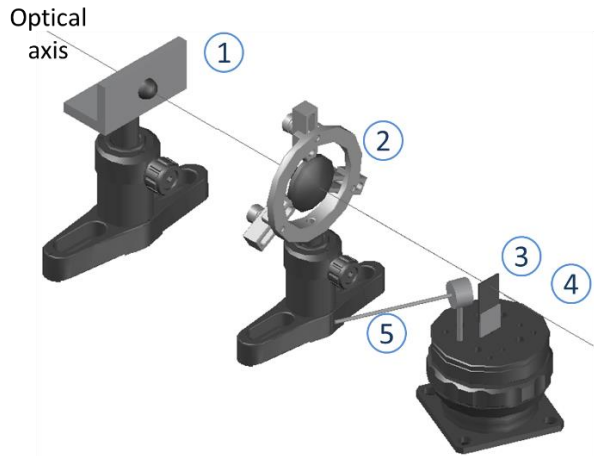

1. LED.
2. Lens.
3. Sample.
4. Height-adjustable swivel positioner.
5. Optical fiber.

Figure 1. Diagram of the optical setup used to measure photoluminescence.

## 3. Results and discussion

We analyzed the distribution of AuNPs synthesized by the in situ reduction of $HAuCl_4$ in both fresh and partially oxidized p-Si. Fig. 2 shows the cross-sectional SEM images of AuNPs in (a) fresh and (b) partially oxidized p-Si on the c-Si substrate, where the dipping time in both cases was 60 min. These images were obtained using back-scattered electrons (BSE). As can be observed, the AuNPs in fresh p-Si were mainly located on the surface, whereas only a few of them were deep inside the porous layer. This effect could be the result of fast reduction of gold ions by H on the p-Si surface [17]. In contrast, AuNPs were well distributed inside the p-Si layer that was partially oxidized prior to dipping, and it was observed that in this case, AuNPs tended to aggregate. In this case, gold ions were reduced at the reactive sites of partially oxidized p-Si, with the remaining H on the surface area of p-Si [24]. Fig. 3 shows the corresponding PL spectra of the two samples. The PL intensity of p-Si partially oxidized with AuNPs was more than four times higher than that of the fresh p-Si. As expected, the PL enhancement due to the surface plasmon effect was higher when AuNPs were well distributed in the porous structure of p-Si than on the surface. The PL spectra were deconvoluted by fitting four Gaussian curves (inset of Fig. 3). The PL spectrum of fresh p-Si with AuNPs consisted of four Gaussian bands (Table 2) with the highest intensity observed at 541 nm, and a full width at half maximum (FWHM) of 105 nm. All bands could be due to quantum confinement in p-Si and possibly coupled to the surface plasmon of AuNPs [25]. However, their intensities were low, possibly because of the high absorption of AuNPs. In contrast, p-Si partially oxidized with AuNPs has four bands; the band at 656 nm is the most intense and widest, which corresponds to the emission of oxidized p-Si, as has been extensively reported previously [26]. The bands at shorter wavelengths are a product of quantum confinement in p-Si. All bands were enhanced by the surface plasmon effect of the AuNPs. It is possible that there are several bands owing to the different sizes of AuNPs. Therefore, we continued to study the synthesis of AuNPs in partially oxidized p-Si because of the superior features shown in the results discussed above.

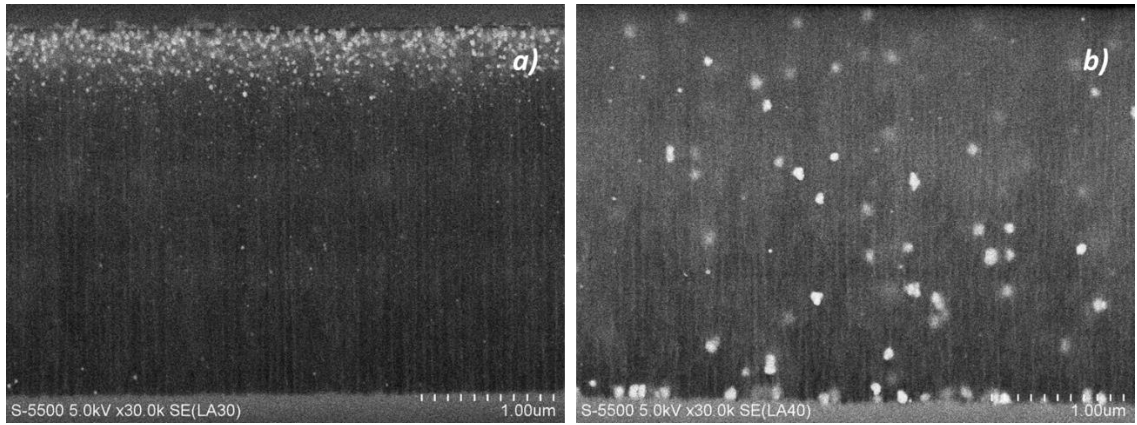

Figure 2. Cross section of *a)* fresh and *b)* oxidized p-Si layers with AuNPs.

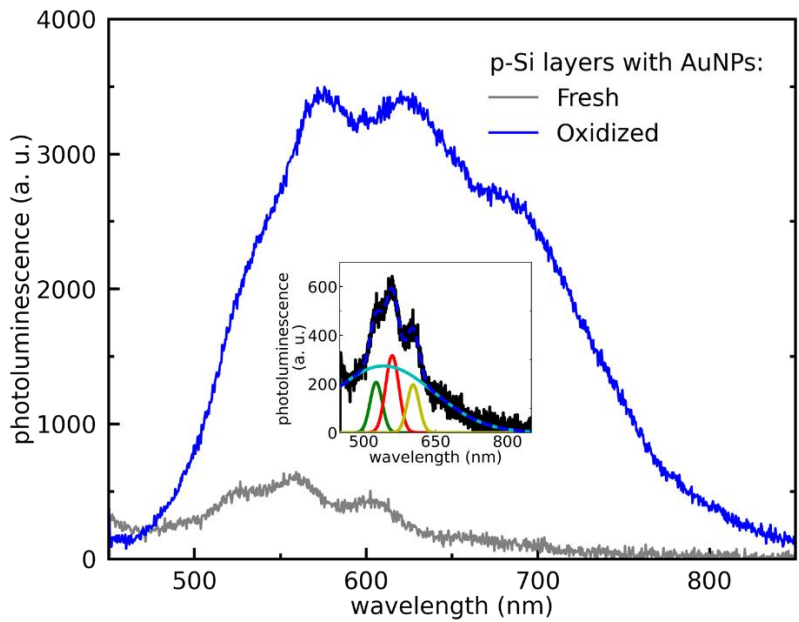

Figure 3. Photoluminescence spectra of AuNPs in *a)* fresh and *b)* oxidized p-Si.

The partially oxidized p-Si samples were immersed in the HAuCl$_4$ solution for different periods of 30, 60, 90, and 120 min. Fig. 4 shows the cross-section (a) and surface images (b) of the p-Si layer with AuNPs synthesized for 90 min. AuNPs of different sizes, up to approximately 100 nm, were formed in the p-Si layer, and few of them were synthesized on the surface, allowing the HAuCl$_4$ to penetrate the pores. The size of the AuNPs increases from the surface (top) to the depth (bottom) of the p-Si layer.

The elemental weight percentages of the p-Si layers with AuNPs were obtained by EDS analysis, and the results are shown in Table 1. The samples contained Si, O and Au atoms. The first two correspond to the structures of the oxidized p-Si and the substrate. The weight percentage of Au in the samples increased with dipping time, but from 90 to 120 min, the weight percentage of Au remained constant, possibly because of the depletion of HAuCl$_4$ in the solution.

Table 1. Elemental weight percentage of the simple p-Si layers with AuNPs obtained by EDS

| Dipping time in chloroauric acid solution (min) | Elements (% weight) | | |
|---|---|---|---|
| | O | Si | Au |
| 30 | 36.59 | 58.16 | 5.25 |
| 60 | 38.24 | 52.74 | 9.02 |
| 90 | 37.47 | 47.84 | 14.69 |
| 120 | 37.33 | 48.31 | 14.36 |

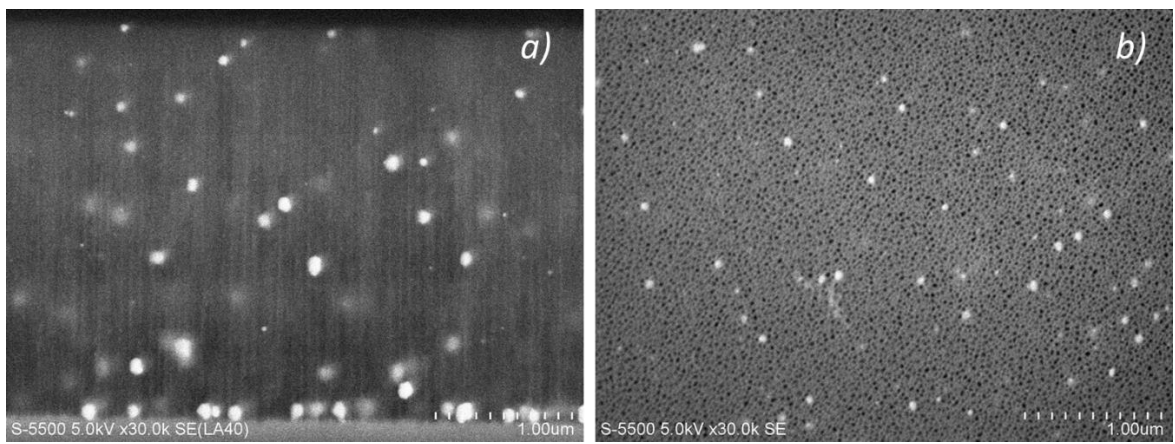

Figure 4. *a)* Cross-sectional and *b)* surface microscopy images of partially oxidized p-Si with AuNPs synthesized for 90 minutes.

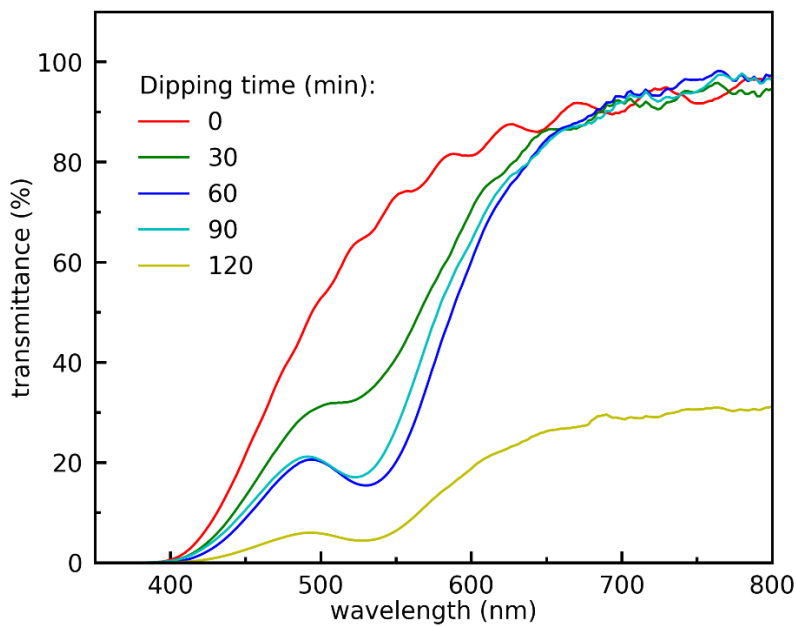

Figure 5. Transmittance of the p-Si layers detached from the substrate with and without AuNPs.

Fig. 5 shows the transmittance spectra of the p-Si layers detached from the c-Si substrate, with and without AuNPs, for dipping times of 30, 60, 90, and 120 min. The transmittance spectrum of the p-Si layer without AuNPs shows the typical interference of Fabry–Pérot fringes, and the intensity rapidly decreases to zero at approximately 400 nm owing to the strong p-Si absorption in that region. P-Si with AuNPs showed a valley around 526 nm in the transmittance spectra owing to the plasmon absorption of AuNPs with a corresponding average particle size of 83 nm. Absorbance was calculated using the Mie model with the particle size and refractive index of p-Si [27]. The fitting was obtained by matching the maximum theoretical absorbance spectrum to the minimum plasmon transmittance valley. Furthermore, the refractive index was indirectly calculated from the porosity (90%) using the symmetric Bruggeman effective medium approximation, as described by Estrada-Wise and del Río [28]. The absorption of the AuNPs increased from dipping times of 30 min to 90 min because of the increase in the number of synthesized AuNPs. The high concentration of AuNPs produced with a dipping time of 120 min increased the absorption of the material at all wavelengths, and attenuated the transmittance; however, plasmon absorption was still observed.

Fig. 6 shows the FTIR spectra of the p-Si samples on their c-Si substrates: as-prepared, oxidized with hydrogen peroxide solution, and with AuNPs synthesized in situ by dipping in the $HAuCl_4$ solution for 30, 60, 90, and 120 min. The as-prepared p-Si sample shows peaks corresponding to the Si-$H_x$ bonds (x = 2, 1) at 2110 and 2090 cm$^{-1}$. At lower wavenumbers, the $SiH_2$ scissor mode appears slightly shifted to 900 cm$^{-1}$ and the Si-Si bond appears at 615 cm$^{-1}$. The small valley at 1050 cm$^{-1}$ corresponds to the slight oxidation of p-Si upon exposure to air after the electrochemical etching.

The partially oxidized p-Si alone and p-Si with AuNPs samples show all the same peaks, as these last samples were oxidized in the hydrogen peroxide solution before the in situ AuNPs synthesis. The intense broad band centered at approximately 3390 cm$^{-1}$ was related to the -OH species of water and ethanol residuals. The isolated small peak at 3740 cm$^{-1}$ is due to the Si-OH bound stretching, with the bending mode of Si-OH at 1640 cm$^{-1}$ [29-31]. A short peak appeared at 2980 cm$^{-1}$ for the C-H stretching bonds, possibly arising from residual ethanol [32,33]. As expected, after oxidation of the samples, the Si-H$_x$ bonds at 2110 and 2090 cm$^{-1}$ were practically eliminated, the peak at 615 cm$^{-1}$ of the Si-Si bond was significantly reduced, and the peak at 1050 cm$^{-1}$ corresponding to the Si-O-Si bonds notably increased [29,34,35]. Furthermore, the SiH$_2$ scissor mode at 900 cm$^{-1}$ corresponds to the oxidized hydride deformation mode δ(-O$_y$Si-H$_x$) at 880 cm$^{-1}$ [34]. No trace of chloroauric acid was detected.

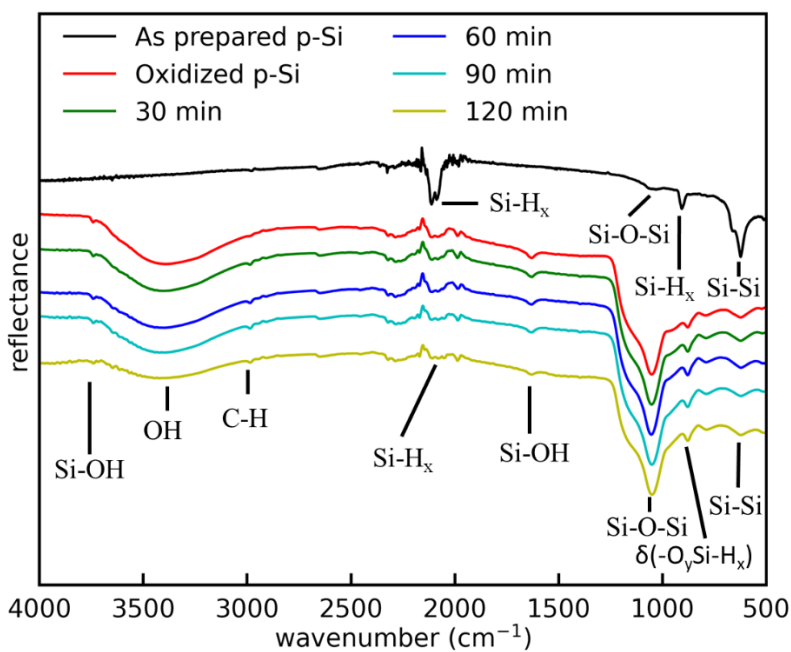

Figure 6. FTIR spectra from p-Si as prepared, partial oxidized and with AuNPs for the dipping times of 30, 60, 90 and 120 min.

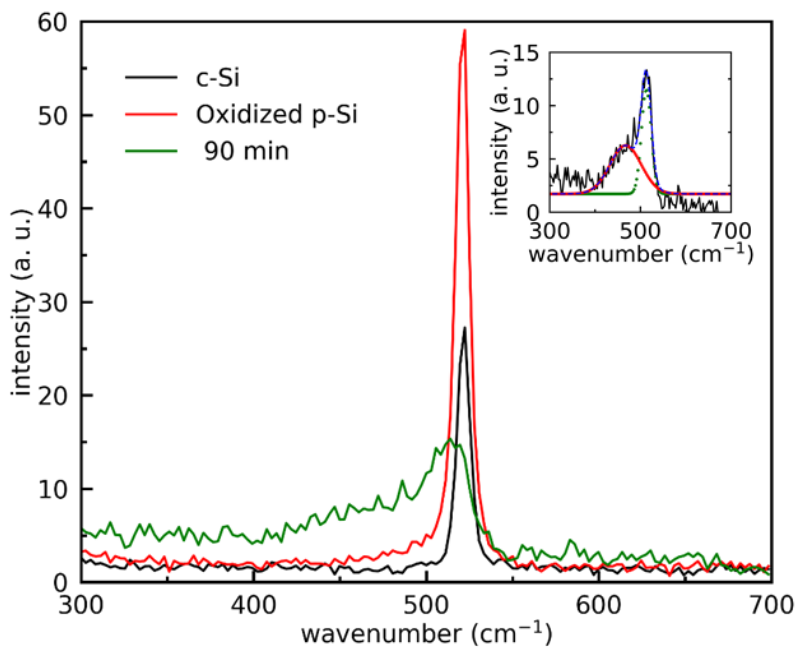

Figure 7. Raman spectra of c-Si substrate, oxidized p-Si layer and p-Si layer with AuNPs of a 90 min dipping time.

Room-temperature Raman spectra of the p-type c-Si substrate used to prepare p-Si (p-type c-Si boron doped with 0.01 Ω cm), partially oxidized p-Si, and p-Si with AuNPs (synthesized during 90 min dipping time) are shown in Fig. 7. The samples were excited with a laser of 532 nm. C-Si and partially oxidized p-Si exhibited a main phonon peak at 521 cm$^{-1}$, with a higher intensity in p-Si owing to the large surface area of the porous structure [36]. The FWHM of c-Si is 8 cm$^{-1}$, whereas the peak for the p-Si layer is slightly wider, with a FWHM of 8.8 cm$^{-1}$, which is again a consequence of the porous structure [37]. The FWHM of the Raman peak at 521 cm$^{-1}$ of the boron-doped c-Si is wider than that of the intrinsic c-Si (4 cm$^{-1}$), because it depends on the temperature, dopant type and laser excitation [38]. In the case of p-Si with AuNPs, this peak widened, reduced in intensity, and shifted compared to that of p-Si alone. Peak deconvolution by fitting Gaussian functions (see inset in Fig. 7) exhibited two peaks at 513 and 464 cm$^{-1}$, with FWHM values of 22 and 83 cm$^{-1}$, respectively. The red-shift of the maximum at 513 cm$^{-1}$ and the new peak at 464 cm-1 are possibly due to the stress caused by the partial oxidation of p-Si [37]. Furthermore, the decrease and widening of the Raman signal are possibly caused by the surface plasmon resonance of AuNPs in p-Si, as observed by Van-The Vo *et al* [39].

Fig. 8 shows the PL spectra of the as-prepared, partially oxidized, and AuNPs-dipped p-Si samples after 30, 60, 90, and 120 min. The applied LED excitation power was 1.30 mW cm$^{-2}$. The spectra were deconvoluted using Gaussian curves to identify different bands (Table 2). The as-prepared p-Si shows weak emission with the most intense band at 658 nm. This band was increased by oxidation and slightly red-shifted. Note that the FWHM of the partially oxidized p-Si band is narrower than that of the as-prepared p-Si band, and it increases for the samples with AuNPs. The PL intensity of the band at 658 nm was red-shifted for the sample partially oxidized for 30 min, and blue-shifted for longer dipping times. This band was also the widest, and its intensity increased for the dipping times of 30 min and 90 min. Note that this band barely appears in the sample with dipping time of 60 min in fresh p-Si (see Fig. 3), possibly because of the high absorption of AuNPs. In particular, the absorption of AuNPs in the sample with a dipping time of 60 min in fresh p-Si was too intense, and its PL was comparable to that of the as-prepared p-Si sample. In addition, the band at approximately 567 nm became largely remarked, which is not the case for fresh p-Si. For this specific emission band, the PL maximum of the sample with an immersion time of 90 min increased six-fold compared to the PL maximum of the as-prepared p-Si.

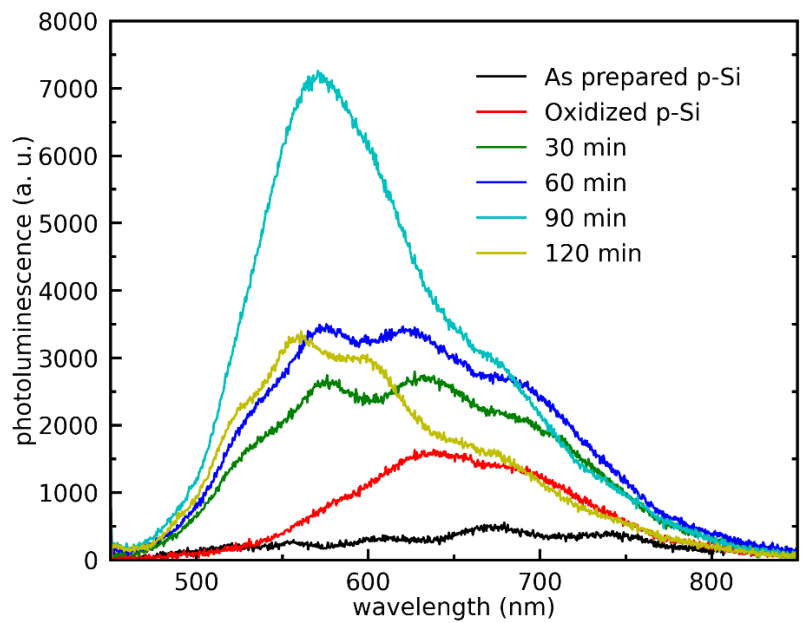

Figure 8. Photoluminescence of simple p-Si layers: as-prepared, oxidized and with AuNPs for dipping times of 30, 60, 90, and 120 min.

Table 2. Band parameters of the samples: as-prepared, oxidized, with AuNPs for dipping times of 30, 60, 90, and 120 min, made in partially oxidized p-Si, and 60 min made in fresh p-Si. The letters "A", "P" and "F" stand for Amplitude (a. u.), Position (nm) and FWHM (nm).

| Sample | Bands | | | | | | | | | | | |
|---|---|---|---|---|---|---|---|---|---|---|---|---|
| | 1 | | | 2 | | | 3 | | | 4 | | |
| | A | P | F | A | P | F | A | P | F | A | P | F |
| As prepared | 298 | 658 | 120 | 144 | 740 | 19 | 190 | 674 | 17 | 47 | 608 | 8 |
| Oxidized | 1437 | 660 | 67 | 128 | 578 | 15 | 265 | 626 | 18 | | | |
| 30 min | 2240 | 662 | 69 | 1017 | 578 | 17 | 690 | 625 | 18 | 1290 | 539 | 29 |
| 60 min | 2839 | 656 | 71 | 1085 | 576 | 16 | 867 | 617 | 18 | 1600 | 541 | 29 |
| 90 min | 3176 | 639 | 70 | 5199 | 567 | 36 | | | | | | |
| 120 min | 1744 | 635 | 74 | 2193 | 554 | 35 | 548 | 606 | 13 | | | |
| 60 min (on fresh) | 196 | 603 | 13 | 316 | 559 | 14 | 274 | 541 | 105 | 207 | 526 | 13 |

The PL spectra of the p-Si layer with AuNPs and a dipping time of 90 min were measured at different excitation powers, from 0.35-1.58 mW cm$^{-2}$, as shown in Fig. 9. As expected, the PL increased with the excitation power. Two Gaussian curves were fitted to the photoluminescence spectra, as shown in the inset of Fig. 10. These curves were centered at approximately 567 nm and 639 nm. Fig. 10 shows the PL maxima of the two fitted Gaussian curves as a function of excitation power. The PL intensity increased linearly, and saturation was not reached.

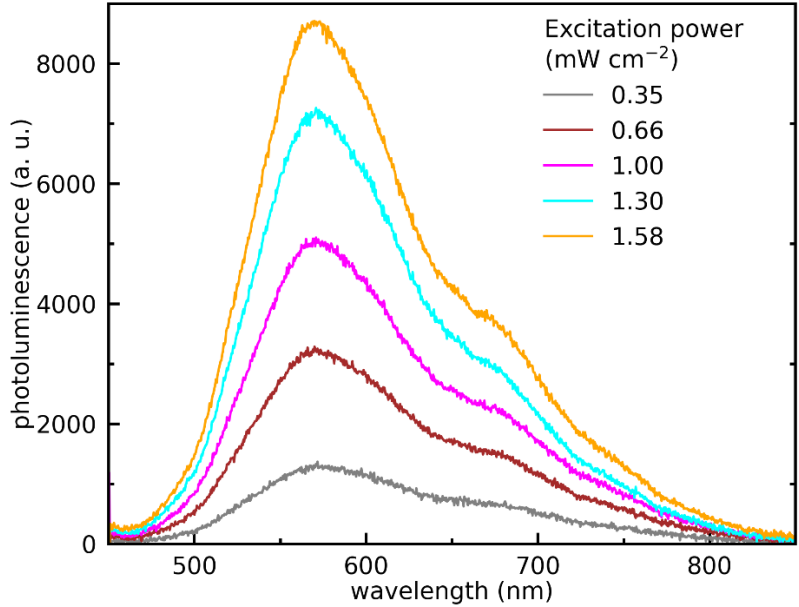

Figure 9. Photoluminescence spectra of the simple p-Si layer with AuNPs and a dipping time of 90 min, the excitation power was varied from 0.35 to 1.58 mWcm$^{-2}$.

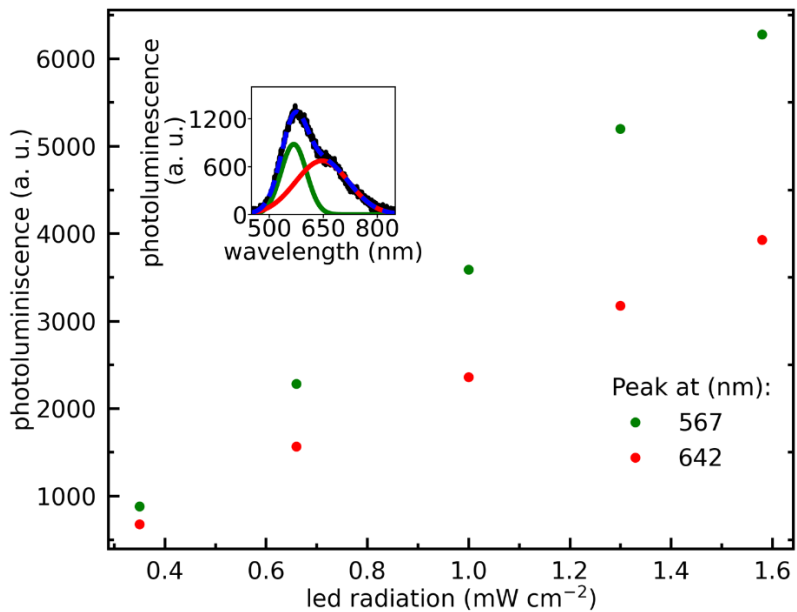

Figure 10. Photoluminescence maxima of the two fitted Gaussian curves as function of excitation power, for the p-Si with AuNPs and a dipping time of 90 min.

To corroborate the enhancement results presented above, excitation spectra were acquired. Fig. 11 shows the PL at 650 nm as a function of the excitation wavelength, from 420 to 600 nm for the p-Si layers: oxidized and with AuNPs for dipping times of 30, 60, 90, and 120 min. A minimum localized at approximately 530 nm was observed for the p-Si layers with AuNPs, which is directly related to the plasmon resonance, as shown in Fig. 5. This result confirms the coupling plasmonic effect, from which the emission band at 567 nm is directly enhanced. Indeed, it is also related to the lack of saturation observed in Fig. 10 for the increased emission with increasing excitation power. Further investigation is required to explore all the potential related to this result, in particular, to study the possibility of achieving laser emission from this p-Si photonic structure.

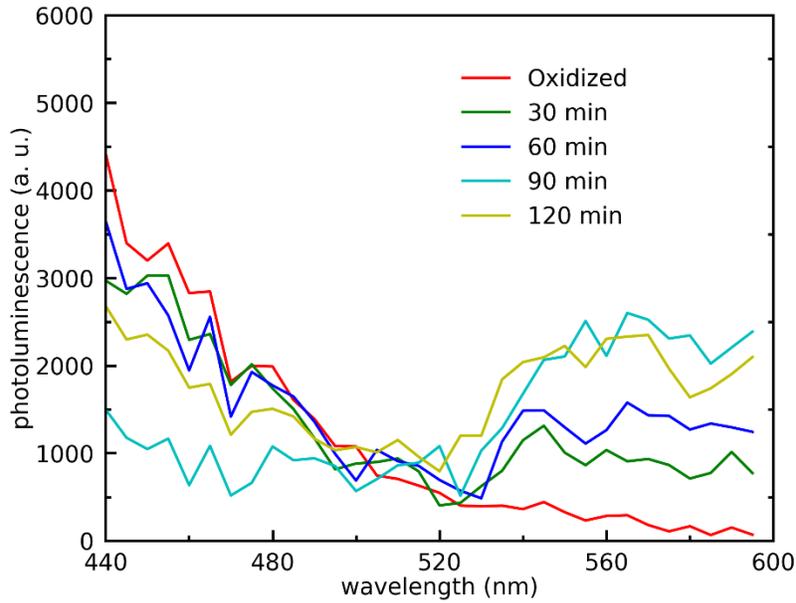

Figure 11. Excitation spectra of p-Si layers: oxidized and with AuNPs for dipping times of 30, 60, 90 and 120 min, measured at an emission of 650 nm.

## 4. Conclusions

We fabricated porous silicon with embedded AuNPs by the in situ reduction of chloroauric acid. In the fresh p-Si samples, AuNPs grew mainly on the surface of the p-Si layer. In contrast, the AuNPs were uniformly distributed inside the partially oxidized porous silicon. AuNPs aggregates of different sizes were obtained, with an average size of 63 nm, corresponding to a plasmon resonance of approximately 526 nm. According to the Au content, the number of AuNPs increased with dipping time up to a maximum of 14.69 %. As a result, a six-fold enhancement of the PL from the emission band at 567 nm, and a linear increase in the PL intensity with an increase in the excitation power were observed, and no saturation was reached. Active 1D photonic structures can be produced with this type of p-Si [26], opening the possibility of controlling and enhancing light emission in optoelectronic devices.

**Acknowledgments**

This research was partially funded by the ECOS-Nord CONAHCyT-ANUIES 315658, PAPIIT-UNAM IN112022, and PAPIIT-UNAM IN109122. J.A.R.E. is grateful for the sabbatical funding from PASPA-UNAM, CONAHCyT, and the University of Sherbrooke. S.E.G. thanks CONAHCyT for postdoctoral fellowship. The authors wish to acknowledge the technical assistance provided by Gerardo Daniel Rayo López, and José Campos Álvarez for the SEM images and EDS measurements.